\documentclass[10pt, a4paper]{article}

\usepackage{amsmath}
\usepackage[dvipsnames]{xcolor}
\usepackage{amscd}
\usepackage{amsthm}
\usepackage{amssymb}
\usepackage{amsfonts}
\usepackage{amsrefs}
\usepackage{algorithm}
\usepackage{algpseudocode}
\usepackage{enumerate}
\usepackage{bm}
\usepackage{mathtools}
\usepackage{mathdots}
\usepackage[T1]{fontenc}
\usepackage{framed, color}
\usepackage{dsfont}
\usepackage{graphicx}
\usepackage{hyperref}
\usepackage{latexsym}
\usepackage{mathrsfs}
\usepackage{subcaption}
\usepackage{verbatim}
\usepackage{booktabs} 
\usepackage{placeins}
\usepackage{thmtools, thm-restate}
\usepackage{multirow}
\usepackage{todonotes}
\usepackage{dsfont}

\theoremstyle{plain}
\newtheorem{theorem}{Theorem}[section]

\newtheorem{remark}[theorem]{Remark}

\theoremstyle{definition}
\newtheorem{definition}[theorem]{Definition}

\newcommand{\RR}{\mathbb R}
\newcommand{\CC}{\mathbb C}
\newcommand{\NN}{\mathbb N}
\newcommand{\PP}{\mathbb{P}}
\newcommand{\EE}{\mathbb{E}}
\newcommand{\FF}{\mathbb{F}}

\title{Model-based Dynamic $3$D MRI Reconstructions using Neural Fields and Tensor Product Expansions}

\author{Ray Sheombarsing\thanks{
All authors are affiliated with the Computational Imaging Group for MRI Therapy \& Diagnostics,
Department of Radiotherapy, University Medical Center Utrecht, Utrecht, The Netherlands.}, Max van
Riel, David Heesterbeek, \\ Nico van den Berg, Alessandro Sbrizzi}

\begin{document}

\maketitle

\begin{abstract}
    Conventional MRI reconstruction methods treat images and coil sensitivities as discrete objects,
    leading to high memory demands and limited structural awareness that hamper effective
    regularization. These limitations hinder accurate reconstruction in highly undersampled
    scenarios, such as dynamic 3D cardiac magnetic resonance (CMR). We introduce a
    discretization-free, memory-efficient, model-based framework for dynamic 2D and 3D MRI
    reconstruction from highly undersampled data. We represent magnetization and coil sensitivities
    as continuous objects---differentiable functions---using tensor products of univariate neural
    fields. This tensor product structure enables scalable optimization in high-dimensional
    spatiotemporal settings. Our method outperforms
    state-of-the-art model-based reconstructions in dynamic 2D and 3D MR settings, preserving
    structure and motion even under aggressive undersampling (e.g., acceleration factor 16). 
\end{abstract}

\begin{center}
{\bf \small Keywords} \\ \vspace{.05cm}
{ \small Accelerated MRI $\cdot$ Continuous-domain modeling $\cdot$ Dynamic 3D MRI $\cdot$ Joint
image-sensitivity reconstruction $\cdot$ Neural fields $\cdot$ Tensor product representations}
\end{center}

\section{Introduction}
Conventional MRI reconstruction represents magnetization (images) as discrete arrays. While natural
for digital data, this approach suffers from high memory demands and limited structural awareness,
requiring strong regularization (e.g., Total Variation) to enforce consistency, especially in highly
undersampled acquisitions such as dynamic $2$D and $3$D cardiac magnetic resonance (CMR)
\cite{tv1,tv2,tv3,arshad,olausson2025free,otazo2015low,biswas2019dynamic}. In these applications,
high acceleration factors are essential to resolve cardiac motion at sufficiently fine
spatiotemporal resolution. Under such conditions, the limitations of purely discrete model-based
frameworks become evident, as effective regularization becomes increasingly difficult.

Deep learning–based reconstruction methods \cite{cine1,cine2,cine3} provide an alternative by
learning expressive priors from large collections of fully sampled data. Broadly, these approaches
fall into two categories. Fully supervised approaches use neural networks to map undersampled data,
or aliased images, to fully sampled $k$-space or directly to images. Physics-informed model-based
approaches incorporate deep networks into classical parallel imaging, typically replacing
handcrafted regularizers with convolutional neural networks trained on large paired datasets.
Despite their success, these methods share two limitations: $(i)$ they rely heavily on large
supervised training sets, reducing adaptability across scanners, protocols and region of interests, 
and $(ii)$ they treat images as discrete objects, tying reconstructions to a fixed resolution and
limiting structural flexibility.

These limitations motivate scan‑specific continuous representations, where images (and coil
sensitivities) are modeled as differentiable functions of space and time. Continuous representations
offer several advantages: they support evaluation and differentiation at arbitrary points, enable
discretization‑free regularization, and connect naturally to well‑established tools in scientific
computing such as orthogonal polynomial expansions \cite{cheney1998introduction}, splines
\cite{schumaker2007spline,de1978practical}, and wavelets
\cite{walnut2013introduction,pereyra2012harmonic}. Importantly, multivariate functions can be
represented efficiently using tensor products of univariate functions, enabling fast evaluation and
differentiation.

Directly applying these classical tools to MRI reconstruction, however, requires choosing a specific
functional representation, e.g., a basis. In highly undersampled scenarios, this choice effectively
acts as a strong prior, which may limit performance. Implicit Neural Fields (INFs)
\cite{molaei2023implicit,SIREN} offer a flexible alternative: they represent functions with neural
networks that adapt to the structure of each scan. Like classical representations, INFs yield
continuous representations, but in contrast, the entire representational structure is learned during
reconstruction rather than fixed in advance. Existing INF‑based MRI approaches mostly employ fully
multivariate networks \cite{spatiotemporal_INF,imjense,nerp,pisco,molaei2023implicit}. Because the
computational cost of these models grows rapidly with dimensionality, they become difficult to apply
in higher‑dimensional (dynamic) problems.

In this paper, we propose a model‑based reconstruction framework using tensor products of univariate
neural fields for dynamic $2$D and $3$D MRI. This architecture generalizes classical tensor product
bases, such as Fourier and polynomial bases, replacing fixed basis functions with adaptable neural
networks. These networks provide continuous, scan‑specific representations of both magnetization and
coil sensitivities, which are reconstructed jointly from undersampled data without requiring
external training datasets or separate coil‑calibration scans.

Our main contribution is an efficient formulation that uses tensor products of univariate neural
fields rather than a single multivariate network. This structure is crucial for computational
feasibility; evaluating a multivariate neural field on a dense $P_{1} \times \ldots \times P_{d}$
grid involves $\prod_{j=1}^{d} P_{j}$ points to evaluate at, which is prohibitive for dynamic $3$D
MRI ($d=4$). In contrast, our tensor product design evaluates only $d$ univariate networks at
$P_{j}$ points each, drastically reducing computation to $\sum_{j=1}^{d} P_{j}$ evaluations thereby
saving memory. Differentiation is also far more memory‑efficient under the tensor product
formulation, which is crucial when evaluating gradient‑based regularizers such as Total Variation
(TV). Combined with a stochastic formulation of the parallel‑imaging objective, this enables
efficient reconstruction at high acceleration factors. 

To illustrate the practical impact of these memory savings, consider a dynamic $2$D example from our
experiments (Section \ref{sec:numerics}), which uses $d=3$ and grid size $(P_{1}, P_{2}, P_{3}) =
(8, 288, 112)$. A conventional multivariate neural field requires roughly $23$GB GPU memory during
optimization, whereas our tensor product formulation reduces this to about $750$MB, a $31\times$
reduction. The gap widens further in our dynamic $3$D examples. With $(P_{1}, P_{2}, P_{3}, P_{4}) =
(16, 128, 64, 64)$, optimization with a single multivariate network cannot be performed on a $48$GB
GPU, while the proposed method converges successfully using only $12$GB, with additional reductions
possible by using smaller batch sizes in the stochastic optimization.

\paragraph{Related work} 
Implicit neural fields have recently been explored for model‑based MRI reconstruction
\cite{spatiotemporal_INF,imjense,nerp,pisco} and, more broadly, for inverse problems. As in
conventional reconstruction, existing approaches can be grouped into methods operating in frequency
space (GRAPPA‑like \cite{grappa}) and those in the spatial domain (SENSE‑like \cite{sense}). Since
our formulation belongs to the latter class, we focus on spatial‑domain approaches.

A closely related approach is IMJENSE \cite{imjense}, which represents magnetization using a neural
field while modeling coil sensitivities with low‑order polynomials. This design choice reflects the
smoothness of coil maps, and as the authors argue, avoids high parameter counts which may make the
problem too ill-posed. IMJENSE uses SIREN networks \cite{SIREN}, as we do, but without a tensor
product structure, and its evaluation is restricted to static 2D imaging.

A more recent method \cite{spatiotemporal_INF}, extends neural fields to dynamic 2D MRI. In contrast
to IMJENSE, coil sensitivities are treated as precomputed discrete arrays rather than learned
continuous functions. High‑frequency details are captured using hash encoding \cite{hash}, which
enables standard ReLU‑MLPs to approximate fine structure. However, regularization remains tied to
discrete formulations (e.g., low‑rank constraints on Casorati matrices), and the method is not
applied to higher‑dimensional dynamic data.

Our work differs in several fundamental aspects; the primary contribution is the introduction of a
tensor product architecture of univariate neural fields. This structure replaces a single
multivariate neural field and enables memory‑ and flop‑efficient evaluation, differentiation, and
integration---capabilities that are crucial for dynamic 3D MRI. In addition to this core
contribution, we represent both magnetization and coil sensitivities as neural fields and
reconstruct them jointly. To address the ill‑posedness of joint estimation, we introduce principled
regularization---including (piecewise) smoothness penalties---for both components. Finally, we
formulate the reconstruction objective stochastically, allowing optimization via stochastic gradient
descent and offering fine‑grained control over memory usage. Taken together, these design elements
drastically reduce the memory footprint and make dynamic 3D MRI reconstruction feasible on standard
GPU hardware.

We remark that tensor product decompositions of neural fields have been studied in the broader
neural field literature \cite{original_tensor_products}, as have techniques for designing activation
functions and capturing high‑frequency content (e.g., hash encodings \cite{hash}). Hash encodings
provide a complementary route for efficient multivariate neural field evaluation by learning
multi‑resolution encodings of dense grids and performing fast lookups to map raw coordinates. Our
approach---based on tensor products---is an alternative strategy that achieves efficiency through
structural decomposition rather than learned encodings. In this work, we focus on demonstrating that
tensor product architectures, combined with stochastic model‑based optimization, enable
computationally feasible and high‑quality dynamic 3D MRI reconstruction.

\paragraph{Overview}
This paper is organized as follows. In Section \ref{sec:neural_field_expansions} we introduce the
notion of a neural field expansion using tensor products of univariate multi layer perceptrons
(MLPs). In Section \ref{sec:parallel_imaging} we use these expansions to represent the magnetization
and coil sensitivities as (continously) differentiable functions, and set up the ``continuous''
analog of the classical ``discrete'' parallel imaging problem. We explain how to solve the resulting
model-based reconstruction problem in a memory-efficient way using stochastic gradient descent
(SGD). In Section \ref{sec:numerics} we validate our methodology by applying it to dynamic $2$D and
$3$D MRI reconstruction. An open-source implementation of the proposed method is available at
\url{https://github.com/RaySheombarsing/nfrecon}.

\section{Neural fields and tensor product expansions}
\label{sec:neural_field_expansions}
In this section we introduce the notion of a Neural Field Expansion (NFE). These expansions will be
used to model the magnetization and coil sensitivities as continuously differentiable mappings. The
main idea of an NFE is inspired by the notion of a generalized Fourier series, which we briefly
review first. Afterwards, we make the analogy explicit and define the notion of an NFE. 

\subsection{Neural field expansions} Consider a $d$-dimensional rectangular domain 
$\mathcal{R}_{d} := \prod_{j=1}^{d} [a_{j}, b_{j}]$, where $d \in \NN$ and $a_{j} < b_{j}$ for $1
\leq j \leq d$. In our applications, $\mathcal{R}_{d}$ will be the spatiotemporal domain on which
the magnetization and coil sensitivites are defined. A convenient way to represent a large class of
functions on $\mathcal{R}_{d}$ is to construct an orthogonal basis $\left \{ \Psi_{k}: k \in
\NN_{0}^{d}  \right \}$ for $L^{2}(\mathcal{R}_{d})$. In this case, for any $f \in
L^{2}(\mathcal{R}_{d})$, there exist unique coefficients $c \in \ell^{2}(\NN_{0}^{d})$ such that $
f(x) = \sum_{k \in \NN_{0}^{d}} c_{k} \Psi_{k}. $ Such an expansion is commonly referred to as a
generalized Fourier series. The basis coefficients $c$ are often referred to as the generalized
Fourier coefficients.

\begin{figure}[tb!]
	\centering
    	\includegraphics[width=0.7\linewidth,clip,trim=0 0 0 0]{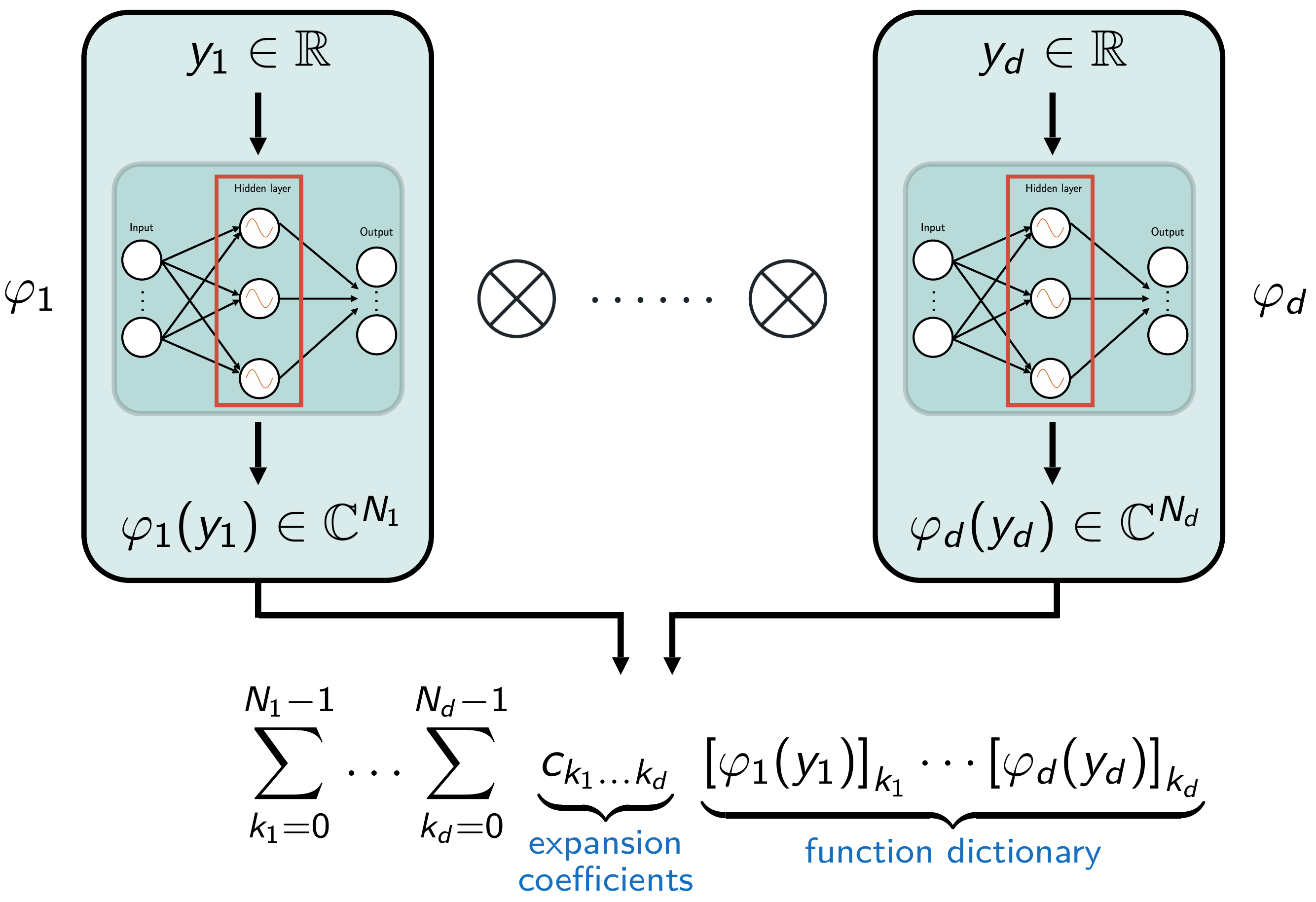}
    	\caption{Schematic overview of a Neural Field Expansion (NFE). To each coordinate 
	direction, we associate a univariate MLP $\varphi_{j}$, where $1 \leq j \leq d$.  
	The components of $\varphi_{j}$ are scalar univariate functions, which we interpret
	as learnable analogs of Fourier modes. 
	We define a subspace of $d$-variate mappings by taking the span of the tensor products of the
	components of $\varphi_{1}$, \ldots, $\varphi_{d}$. Elements in this subspace are referred to as NFEs.}
    	\label{fig:nfe}
\end{figure}

A popular choice in scientific computing, and mathematics in general, for constructing orthogonal
bases on higher dimensional domains is to take the tensor product of one dimensional bases. More
precisely, suppose $\left( \psi_{jk_{j}} \right)_{k_{j} \in \NN_{0}}$ is an orthogonal basis for
$L^{2}([a_{j}, b_{j}])$, then taking the tensor product over the different coordinate directions, we
obtain the orthogonal basis $\left \{ \Psi_{k}: k \in \NN_{0}^{d} \right \}$ for
$\bigotimes_{j=1}^{d} L^{2}\left( [a_{j}, b_{j}] \right) \simeq L^{2} \left( \mathcal{R}_{d}
\right)$, where 
\begin{align}
    \label{eq:orthogonal_basis}
    \Psi_{k}= \psi_{1k_{1}} \otimes \ldots \otimes \psi_{dk_{d}}, \quad k \in \NN_{0}^{d}.
\end{align}
We will use this tensor product construction, at least formally, to define an NFE and represent
mappings in $L^{2}(\mathcal{R}_{d})$. The idea is to simply replace the univariate basis functions
in \eqref{eq:orthogonal_basis} with a \emph{finite} set of univariate MLPs. 
\begin{definition}[Neural field expansion]
	\label{def:nfe}
    Let $N \in \mathbb{N}^{d}$ denote the number of modes for each coordinate direction and
    $\mathbb{F} \in \{\CC, \RR \}$. For each $1 \leq j \leq d$, let $\varphi_{j}: \left [a_{j},
    b_{j} \right] \rightarrow \mathbb{F}^{N_{j}}$ be a MLP with $L \in \NN$ hidden layers, $K \in
    \NN$ hidden units per layer, and activation function $\eta$. The neural field expansion
    associated to $ (\varphi_{1}, \ldots, \varphi_{d})$ with coefficients $c \in \FF^{N_{1} \times
    \ldots \times N_{d}}$ is the map $\Phi: \mathcal{R}_{d} \rightarrow \FF$ defined by 
	\begin{align}
		\label{eq:neural_expansion}
		\Phi(y) := \sum_{k_{1}=0}^{N_{1} - 1} \ldots \sum_{k_{d}=0}^{N_{d} - 1} c_{k_{1} \ldots k_{d}} 
	\left[ \varphi_{1}(y_{1}) \right]_{k_{1}} \cdot \cdot \cdot \left[ \varphi_{d}(y_{d}) \right]_{k_{d}}.
	\end{align}	
\end{definition}
\begin{remark}
    For brevity, we will frequently refer to $(L,K) \in \NN^{2}$ as the architecture of the
    underlying MLPs. Here we have chosen to use the same architecture and activation function for
    each coordinate direction, but these could in principle be different. 
\end{remark}
\begin{remark}
	In this paper we are exclusively concerned with complex-valued mappings. For this reason
	we henceforth take $\FF = \CC$. 
\end{remark}
Each component of $\varphi_{j}$, \emph{a univariate function}, may be thought of as a learnable
analog of a Fourier mode; they are the analogs of the univariate basis functions $\psi_{jk_{j}}$.
Similarly, $c \in \CC^{N_{1} \times \ldots N_{d}}$ is the analog of the ``generalized Fourier
coefficients''. See Figure \ref{fig:nfe} for a schematic overview. The optimizable parameters in an
NFE are $c$ and the learnable parameters $\Theta \in \RR^{p}$ of the neural networks $(\varphi_{1},
\ldots, \varphi_{d})$. In particular, an NFE is fully specified by the parameters
$(L,K,N,\eta,\Theta, c)$, where $(L, K, N, \eta)$ are hyperparameters, and $(\Theta,c)$ the
learnable parameters. To draw the analogy with \eqref{eq:orthogonal_basis}, note that we have
formally replaced $\psi_{jk_{j}}$ with $[\varphi_{j}]_{k_{j}}$ and $\Psi_{k}$ with 
\begin{align}
    \label{eq:neural_field_basis} [\varphi_{1}]_{k_{1}} \otimes \ldots \otimes
    [\varphi_{d}]_{k_{d}}, \quad 0 \leq k_{j} \leq N_{j} -1, \ 1 \leq j \leq d. 
\end{align}
There is one crucial difference, however; the mappings in \eqref{eq:neural_field_basis} are in
general not orthogonal and do not constitute a basis for $L^{2}(\mathcal{R}_{d})$. Nonetheless, the
Universal Approximation Theorem (UAT) still holds in the following sense.

\begin{theorem}[Universal approximation theorem for NFEs]
    \label{thm:UAT}
    Let $\eta$ be a non-polynomial activation function. Then for any $\varepsilon >0$ and any $f \in
    L^{2}(\mathcal{R}_{d})$, there exists a neural field expansion $\Phi \in L^{2}(\mathcal{R}_{d})$
    with parameters $(L,K,N,\Theta,c) \in \NN \times \NN \times \NN^{d} \times \RR^{p} \times
    \CC^{N_{1} \times \ldots \times N_{d}}$ such that $\Vert f - \Phi
    \Vert_{L^{2}(\mathcal{R}_{n})} < \varepsilon$.
\end{theorem}

The proof is relatively straightforward; a slightly different formulation of this theorem can be
found in \cite[Theorem $1$]{original_tensor_products}, whose proof is essentially the same.
Furthermore, the requirement that $\eta$ is non-polynomial can be relaxed, see
\cite{Kidger2019UniversalAW} for instance. Carefully note, however, that Theorem \ref{thm:UAT},
essentially the standard UAT, is only a statement about NFEs being dense in
$L^{2}(\mathcal{R}_{d})$. Ideally, to really have an analogy with generalized Fourier expansions, we
would like to have a more ``constructive'' statement, where we \emph{fix} $(L,K)$, and get to
approximate any $f$ arbitrarily well as we increase the number of modes $N$, while leaving the lower
order modes untouched. The analysis of such a statement, however, is much more involved, and not the
focus of this paper. Here we shall be content with the knowledge that any map $f \in
L^{2}(\mathcal{R}_{d})$, e.g., the magnetization or coil maps, can in principle be approximated to
any desired precision by a well-chosen NFE. 

Let us also note that without imposing orthonormality, and not having linear independence in
particular, the ``importance'' of a mode, cannot be assessed by simply analyzing the magnitude of
the associated coefficient. These observations pose no obstructions for the effectiveness of our
method, but they do obstruct the interpretability of our learned representations. 

\subsection{Efficient evaluation, differentiation and integration}
An important feature of NFEs, and tensor product expansions in general, is that they can be very
efficiently evaluated, differentiated and integrated. These computations effectively reduce to $1$D
computations. More precisely, if $\left \{ \mathcal{P}_{j} \subset [a_{j}, b_{j}]: 1 \leq j \leq d
\right \}$ are finite subsets, then \eqref{eq:neural_expansion} can be evaluated on $\prod_{j=1}^{d}
\mathcal{P}_{j}$ by first evaluating each $\varphi_{j}$ separately on $\mathcal{P}_{j}$, and then
combining the results using $c$. In total, this involves only $\sum_{j=1}^{d} \left \vert
\mathcal{P}_{j} \right \vert$ points to evaluate at; $\vert \mathcal{P}_{j} \vert$ points per
network. In contrast, without the tensor product structure, e.g., if we had used a single
multivariate MLP, we would have needed $\prod_{j=1}^{d} \left \vert \mathcal{P}_{j} \right \vert$
evaluations.

NFEs can also be efficiently differentiated. For instance, $\dfrac{ \partial \Phi}{ \partial y_{j}}$
can be computed by only differentiating $\varphi_{j}$ using Automatic Differentiation (AD). The
resulting sum can then be evaluated on grids as before. Similarly, \eqref{eq:neural_expansion} can
be integrated over $\mathcal{R}_{d}$, or other rectangular subdomains, by evaluating $1$D integrals
over the intervals $[a_{j}, b_{j}]$ using any quadrature rule, e.g., Gauss-Legendre.

Altogether, the memory savings are substantial: a $31 \times$ reduction in our dynamic $2$D example, 
and in the dynamic 3D cases a reduction from more than $48$GB to only $9$-$12$GB.

\section{Model-based reconstruction framework for parallel imaging}
\label{sec:parallel_imaging}
In this section we set up a continuous analog of the discrete parallel imaging problem. We model
magnetization and coil sensitivities as continuously differentiable mappings $m: [0, \tau] \times
\mathcal{R}_{n} \rightarrow \CC$ and $S: \mathcal{R}_{n} \rightarrow \CC^{n_{c}}$, respectively,
using NFEs. Here $\tau >0$ is the acquisition duration, $n \in \{2, 3\}$ the spatial dimension,
$\mathcal{R}_{n} := \prod_{j=1}^{n} [-\frac{s_{j}}{2}, \frac{s_{j}}{2}] \subset \RR^{n}$ a
rectangular domain representing the field of view (FOV), and $n_{c} \in \NN$ the number of coils. 

\subsection{NFEs for magnetization and coil sensitivities}
We represent the magnetization as a complex-valued NFE defined over both
time and space, and the coil sensitivities as a complex vector-valued NFE defined over space. To
ensure consistency across coils, we normalize the coil sensitivities along the coil dimension, i.e.,
$S(x) = \tilde S(x) / \left \Vert \tilde S(x) \right \Vert_{2}$, 
where $\tilde S: \mathcal{R}_{n} \rightarrow \CC^{n_{c}}$ is itself a complex vector-valued 
NFE over space. 

Both NFEs are implemented using sinusoidal activation functions. MLP networks with sinusoidal
activation functions are commonly referred to as SIREN networks \cite{SIREN}. The choice for SIREN 
networks is motivated by their demonstrated ability to represent high-frequency signals and their
derivatives with high accuracy. SIREN networks employ so-called frequency scalings---one
applied in the first layer and another in all subsequent layers---to effectively capture
high-frequency content. We will refer to these scalings as frequency embedding parameters.
Specifically, for each coordinate direction, we define a pair of strictly positive numbers
corresponding to these parameters. Furthermore, we adopt the modified weight initialization proposed
in \cite{SIREN} to ensure stable training. 

The parameters of the magnetization field $m$ and the coil sensitivity field $S$ are denoted by $(L,
K, N, \omega, \Theta, a)$ and $(\tilde L, \tilde K, M, \tilde \omega, \Lambda, {\bm{b}})$,
respectively. Here 
\begin{itemize}
    \item $a \in \CC^{N_{1} \times \ldots \times N_{n+1}}$ and $\bm{b} \in \CC^{M_{1} \times \ldots
        \times M_{n} \times n_{c}}$ are the expansion coefficients,
	\item $N \in \NN^{n+1}$ and $M \in \NN^{n}$ the number of modes, 
    \item $(L, K)$ and $(\tilde L, \tilde K) \in \NN^{2}$ specify the architectures of the
        underlying MLPs, 
	\item $\Theta$ and $\Lambda$ are the learnable parameters of the MLPs,
    \item $\omega \in \prod_{j=1}^{n+1}(0, \infty)^{2}$ and $\tilde \omega \in \prod_{j=1}^{n}(0,
        \infty)^{2}$ are the frequency embedding parameters. 
\end{itemize}

\subsection{Parallel imaging problem}
Next, we set up an optimization problem to jointly find the parameters $(\Theta, a)$ and $(\Lambda,
{\bm{b}})$ of the magnetization and coil sensitivities, respectively. The objective consists of a
data consistency term and regularization. The data consistency term penalizes $k$-space deviations
via a weighted $L^{2}$-norm, ensuring uniform treatment across all spatial frequencies.
Regularization includes spatial TV on $m$, a $L^{1}$-penalty on its temporal derivative, and a
squared $L^{2}$-penalty on the derivatives of $S$. In addition, to lower memory requirements and
make computations feasible for dynamic $3$D reconstructions, we introduce a stochastic component by
randomly sampling subsets of coil measurements and spacetime patches during each step of the
optimization process. Below we provide a precise description. 

\paragraph{Data consistency}
We use a weighted $L^{2}$-norm for data consistency rather than the standard squared $L^{2}$-norm.
The latter is usually motivated by maximum‑likelihood estimation under the assumption that the
residuals follow an isotropic Gaussian distribution. Empirically, the weighted $L^{2}$-norm provides
greater stability in the presence of model mismatch and the heavy‑tailed residuals produced by
unbinned readouts and stochastic batching.

Let $\mathcal{C} = \{1, \ldots, n_{c} \}$ and $\mathbb{T} = \{t_{0}, \ldots, t_{I}\} \subset [0,
\tau]$ denote the coil-indices and points in time at which measurements were recorded, respectively.
Let $m_{c} := (t,x) \mapsto m(t,x)S_{c}(x)$ denote the coil-image associated to coil index $c \in
\mathcal{C}$, and $\widehat{m_{c}}$ its spatial Fourier Transform. The data consistency term is
defined by 
\begin{align*}
    \mathcal{L}_{\text{data}}(a, {\bm{b}}, \Theta, \Lambda) &:=  \EE_{\PP_{\mathbb{T}}} \EE_{\PP_{\mathcal{C}}}
    \left(
        (c, t) \mapsto 
        \sqrt{ \sum_{\xi \in \Xi_{t}} \left \vert \widehat{m_{c}}(t, \xi) - \widehat{d_{c}}(t, \xi) \right \vert^{2}
        w_{c}(t, \xi; \epsilon) }
    \right). 
\end{align*}
Here $\Xi_{t} \subset \RR^{n}$ are the spatial frequencies recorded at time $t \in \mathbb{T}$ with
associated $k$-space measurements $\left \{ \widehat{d_{c}}(t, \xi): \xi \in \Xi_{t} \right \}$, and
$\left \{w_{c}(t, \xi; \epsilon): \xi \in \Xi_{t} \right \}$ are weights ensuring that high spatial
frequencies with small magnitude are not ignored. Specifically, given a tolerance $\epsilon > 0$, we
define
\begin{align*}
    w_{c}(t, \xi; \epsilon) := 
    \begin{cases}
        1, &\left \vert \widehat{d_{c}}(t, \xi) \right \vert \leq \epsilon, \\[2ex]
        \left \vert \widehat{d_{c}}(t, \xi) \right \vert^{-\frac{1}{2}}, 
        & \left \vert \widehat{d_{c}}(t, \xi) \right \vert > \epsilon, \\[2ex]
    \end{cases}
\end{align*}
leaving observations with very small magnitude unweighted. Furthermore, $\PP_{\mathbb{T}}$ and
$\PP_{\mathcal{C}}$ are discrete independent uniform probability measures over the recorded times
$\mathbb{T}$ and coil-indices $\mathcal{C}$, respectively.

\paragraph{Regularization of magnetization}
We impose TV regularization on $m$ in both space and time, separately. More precisely, we define 
\begin{align*}
	\mathcal{L}_{\text{TV}x}(a, \Theta) := \EE_{U\left ([0, \tau] \times \mathcal{R}_{n}\right)}
	 \left( 
		\Vert \nabla_{x} m \Vert_{2} 
	\right), \quad
	\mathcal{L}_{\text{TV}t}(a, \Theta) := 
	 \EE_{U\left ([0, \tau] \times \mathcal{R}_{n}\right)}  \left( 
	\left \vert \frac{ \partial m}{\partial t} \right \vert
	\right),	
\end{align*}
where $U\left ([0, \tau] \times \mathcal{R}_{n}\right)$ denotes the (continuous) uniform
distribution on $[0,\tau] \times \mathcal{R}_{n}$. 

\paragraph{Regularization of coil sensitivities}
We penalize the squared $L^{2}$-norm of the partial derivatives of $S$ to promote smoothness. That
is, we define 
\begin{align*}
	\mathcal{L}_{\text{coil}}({\bm{b}}, \Lambda) := 
	\EE_{\PP_{\mathcal{C}}} \EE_{U(\mathcal{R}_{n})} \left(
		(c, x) \mapsto \sum_{j=1}^{n} \left \vert \frac{ \partial S_{c} }{ \partial x_{j}}(x)  \right \vert^{2}
	\right), 
\end{align*}
where $U\left(\mathcal{R}_{n}\right)$ denotes the (continuous) uniform distribution on 
$\mathcal{R}_{n}$.

\paragraph{Complete objective}
Finally, the full objective to be minimized is
\begin{align}
    \label{eq:objective}
	 \mathcal{L}(a, {\bm{b}}, \Theta, \Lambda) :=
	 \mathcal{L}_{\text{data}}(a, {\bm{b}}, \Theta, \Lambda) 
	 + \lambda_{\text{TV}x}\mathcal{L}_{\text{TV}x}(a, \Theta)
	 + \lambda_{\text{TV}t}\mathcal{L}_{\text{TV}t}(a, \Theta)
	 + \lambda_{\text{coil}}\mathcal{L}_{\text{coil}}({\bm{b}}, \Lambda),
\end{align}
where $\lambda_{\text{TV}x}, \lambda_{\text{TV}t}, \lambda_{\text{coil}} > 0$ are weights to control
the strength of the regularization terms.  

\paragraph{Stochastic optimization and discretization}
In practice, optimization is performed using stochastic discretizations of the objective. Although
the regularization terms are defined over the full continuous spatiotemporal and spatial domains,
sampling exclusively from a continuous distribution is suboptimal for stable reconstruction. Data
consistency is imposed only at a discrete set of measurement locations, where reconstruction errors
concentrate and the inverse problem is effectively anchored. Sampling also on the corresponding
Cartesian grid repeatedly constrains these high‑leverage locations, which empirically improves
stability and suppresses localized artifacts. For this reason, at each optimization step we evaluate
the regularization terms on a mixed set of spacetime locations, consisting of both randomly sampled
continuous points and points drawn from the discrete Cartesian grid corresponding to the measurement
locations (using discrete uniform sampling in each coordinate direction for the latter). In
particular, samples are drawn independently in each coordinate direction, which aligns naturally
with the tensor product structure of the employed models and allows the resulting grid‑based
evaluations to be carried out efficiently.

This procedure defines a stochastic, deliberately grid‑biased surrogate of the continuum
regularization functional, which is optimized directly and empirically exhibits substantially
reduced variance and improved stability. 

\begin{figure}[tb]
  \centering
  \begin{subfigure}{0.48\textwidth}
    \centering
    \includegraphics[width=\linewidth,clip,trim=0 0 0 0]{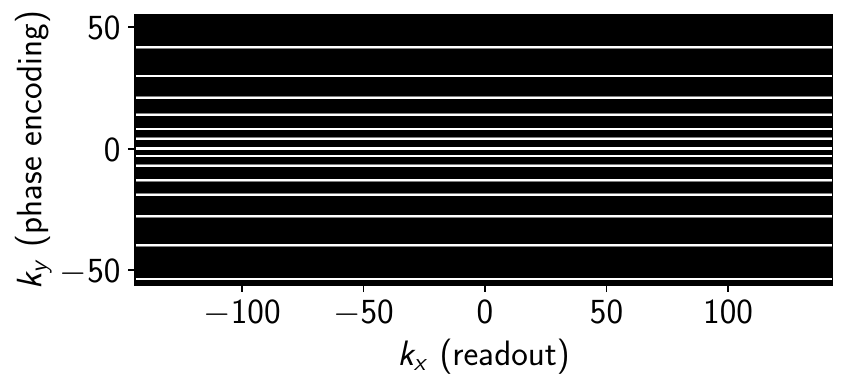}
    \caption{$(i)$ $2$D $+$ time ($\text{AF}=8$).}
    \label{fig:gro_us}
  \end{subfigure} \quad
  \begin{subfigure}{0.48\textwidth}
    \centering
    \includegraphics[width=\linewidth,clip,trim=0 0 0 0]{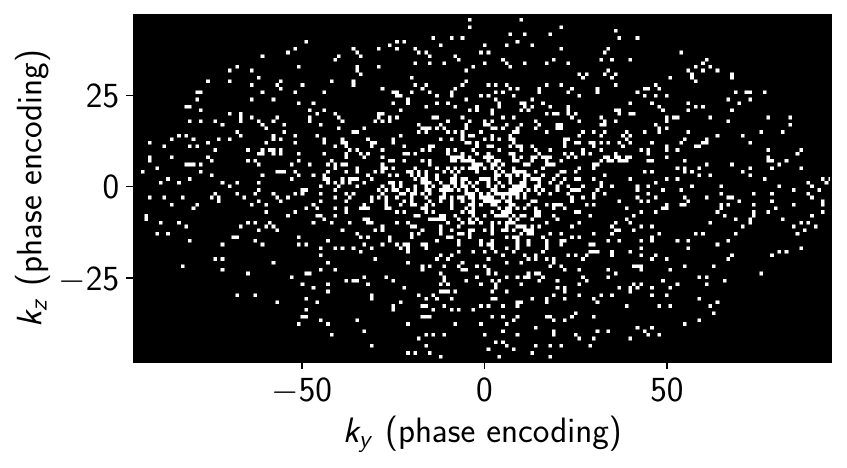}
    \caption{$(iii)$ $3$D $+$ time ($\text{AF}=16$).}
    \label{fig:4d_us}
  \end{subfigure} \\[3ex]
  \caption{
      Cartesian undersampling patterns for (binned) datasets $(i)$ and $(iii)$ at a representative
      time frame. \subref{fig:gro_us} Conventional rectilinear readout pattern at acceleration
      factor $8$. \subref{fig:4d_us} Undersampling pattern at acceleration factor $16$. 
  }
  \label{fig:undersampling_pattern}
\end{figure}

\section{Numerical experiments}
\label{sec:numerics}
In this section we demonstrate the effectiveness of our method by applying it to the reconstruction
of undersampled dynamic $2$D and $3$D MRI data and comparing with state-of-the-art CS
reconstructions.

\subsection{Datasets}
We evaluate our method on three public datasets \cite{ocmr, arshad, Max2026vivo}: $(i)$
quantitatively on retrospectively undersampled dynamic $2$D CMR data, $(ii)$ quantitatively on
prospectively undersampled dynamic $3$D data, $(iii)$ qualitatively by visual inspection on
prospectively undersampled dynamic $3$D CMR data. We quantify image quality in $(i)$ and $(ii)$
using SSIM and PSNR. 
\begin{figure}[tb!]
	\centering
    	\includegraphics[width=0.55\linewidth,clip,trim=0 0 0 0]{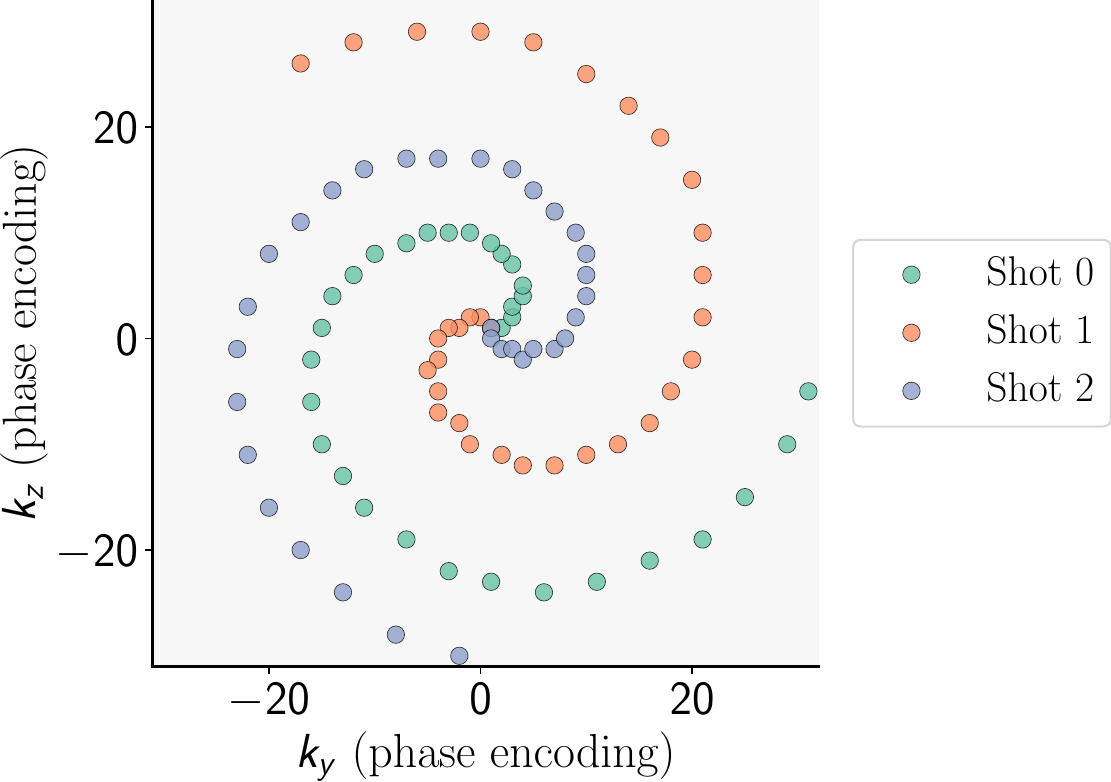}
    	\caption{
            CASPR undersampling pattern for dataset $(ii)$. Each dot corresponds to a single
            readout. The sampling pattern follows a spiral ordering in which every spiral (or
            “shot”) contains $32$ readouts.
        }
    	\label{fig:caspr}
\end{figure}
\paragraph{$(i)$ Dynamic $2$D CMR}
We use fully sampled binned data from the OCMR dataset \cite{ocmr}, consisting of $165$ slices from
$78$ healthy subjects (including all fully sampled $1.5$T and $3$T scans). Ground truth images are
obtained by reconstructing the fully sampled data with coil sensitivity maps estimated via
\textsc{ESPIRIT} \cite{espirit}. Retrospective undersampling is then applied for acceleration
factors $\{8, 12, 16\}$ using the Cartesian rectilinear readout patterns provided in OCMR; see
Figure \ref{fig:gro_us} and \cite{gro}. We reconstruct the undersampled data using our proposed
method and, as a state‑of‑the‑art baseline, also perform CS reconstructions (Section
\ref{sec:cs_details}).

\paragraph{$(ii)$ Dynamic $3$D MRI}
We use the public dynamic $3$D dataset from \cite{Max2026vivo}, which provides both fully sampled
and raw unbinned undersampled data for nine subjects. In both scans, a pressure cuff was placed on
the thigh of a subject, and the pressure was ramped monotonically from zero to a prescribed maximum
and back to zero. In the first scan, the pressure was held fixed at nine predefined levels long
enough to acquire fully sampled Cartesian $k$-space data at each level. In the second scan, the same
pressure cycle was repeated six times without holding at fixed levels. The data in this scan was
prospectively undersampled using a CASPR trajectory \cite{prieto2015highly}. We use the fully
sampled data to obtain ground truth images corresponding to the nine pressure levels, reconstructed
using coil sensitivities from \textsc{ESPIRIT} \cite{espirit}. The proposed method is used to
reconstruct time‑resolved images directly from the raw unbinned data.

\paragraph{$(iii)$ Dynamic $3$D CMR}
We use a prospectively undersampled dynamic $3$D binned CMR dataset (one subject) acquired with a
self-gated free-running $3$D GRE sequence, see \cite{arshad} for the details. The undersampled data
has native acceleration factor $9$, $20$ frames, $28$ receive coils, and an acquisition matrix of
size $115 \times 192 \times 96$.  To further increase undersampling, we retrospectively subsample
each time frame by randomly drawing readouts from a uniform distribution, resulting in an effective
acceleration factor of $16$, see Figure \ref{fig:4d_us}. 

\subsection{Implementation details}
In this section we provide relevant implementation details for performing both the CS and NFE
reconstructions. 

\subsubsection{CS reconstructions}
\label{sec:cs_details}
As state-of-the-art comparisons, we compute CS reconstructions for experiments $(i)$ and $(iii)$ on
the retrospectively undersampled data using low-rank, Total Generalized Variation in time and
wavelet sparsity in space with \textsc{BART} \cite{bart}. For the OCMR dataset $(i)$, where a ground
truth is available, the hyperparameters of the CS reconstruction are tuned via Bayesian optimization
using \textsc{Optuna} \cite{optuna}, maximizing SSIM. Specifically, we perform a Bayesian search for
a specific slice at acceleration rate $8$, and then use those hyperparameters for all the other
slices and acceleration factors. Since no ground truth data is available for the third dataset
$(iii)$, only visual, qualitative comparisons can be performed. Therefore, we select hyperparameters
using a grid search and visual assessment.

We do not perform CS reconstructions for experiment $(ii)$, as this dataset consists of fully
unbinned raw $k$-space readouts that we wish to use directly, without any temporal grouping or
preprocessing. That is, our target temporal resolution is $\delta_{t} = \text{TR}=5.4$ ms. Standard
CS methods operate on binned or frame‑based data and require constructing explicit spatiotemporal
arrays; for unbinned dynamic $3$D data, this would lead to memory demands far beyond practical
limits and is therefore computationally infeasible.

\subsubsection{Proposed NFE reconstructions}
In this section we provide the implementation details for the NFEs. 

\paragraph{Optimization}
\begin{table}
\centering
\scalebox{1.0}{
    \begin{tabular}[tb!]{ccccccc}
        \toprule 
        \textbf{Dataset} &  $b_{\mathcal{C}}$ & $b_{[0, \tau]}$ & $b_{\mathcal{R}_{n}}$ & $\#$ \textbf{Iterations} & \textbf{Learning rate} \\
        \midrule
        $(i)$ & All & $8$ & $(64, 64)$ & $5000$ & $8.0 \cdot 10^{-6}$ \\
         $(ii)$ & $12$ & $16$ & $(48,32,32)$ & $10000$ & $1.0 \cdot 10^{-5}$ \\
         $(iii)$ & $12$ & $8$ & $(64, 64, 32)$ & $3500$ & $5.0 \cdot 10^{-6}$ \\
        \midrule
        \bottomrule				
    \end{tabular}}
    \caption{\label{table:batch_sizes} Batch sizes used during optimization.
        The temporal batch size $b_{[0,\tau]}$ and spatial batch size $b_{\mathcal{R}_n}$ are shown
        per coordinate direction. In all cases, the same number of samples is drawn from the
        continuous domain and from the corresponding discrete partition. We therefore only 
        report one number.}
\end{table}
We minimize the objective in \eqref{eq:objective} using SGD with the weighted \textsc{Adam}
optimizer. The expectations are estimated by independently sampling from the
different coordinate directions (continuous and discrete) and coils using batch sizes
$b_{[0, \tau]}$, $b_{\mathcal{R}_{n}}$ and $b_{\mathcal{C}}$. Here $b_{\mathcal{C}}$ is the number of
coils. The batch sizes $b_{[0, \tau]}$ and  $b_{\mathcal{R}_{n}}$ consist of pairs of integers---one
for the continuous coordinate and one for its associated discrete partition (see Table
\ref{table:batch_sizes}). A learning‑rate scheduler reduces the rate when the data‑consistency term
stagnates for a predefined number of iterations. 

\paragraph{Hyperparameter tuning}
The NFE framework involves many hyperparameters: the architectures for $m$ and $S$, the
per‑coordinate frequency embeddings, regularization weights, learning rate, and the batch sizes.
Without simplifications, this amounts to $2(4n+7)$ parameters (i.e., $30$ for $n=2$, $38$ for
$n=3$). Rather than performing a prohibitively large hyperparameter search for each dataset, we used
the OCMR dataset as a prototypical example to develop practical tuning guidelines. We chose OCMR
because it is the largest and most heterogeneous dataset in our study, both in terms of acquisition
parameters and anatomical coverage.

We first performed an exhaustive Bayesian optimization on a single slice---analogous to the CS
tuning---maximizing SSIM. We then perturbed the resulting hyperparameters, applied them across all
slices, and assessed the mean SSIM to gauge sensitivity. This qualitative analysis yielded several
consistent observations:
\begin{itemize}
	\item[(\textbf{O1})] Increasing the number of modes does not degrade performance.
	\item[(\textbf{O2})] Model capacity can be increased either through more modes or through deeper/wider networks.
	\item[(\textbf{O3})] Coil sensitivities require less model capacity than magnetization.
	\item[(\textbf{O4})] Large initial‑layer frequency embeddings (up to a certain value) do not harm performance; subsequent layers benefit from smaller values.
	\item[(\textbf{O5})] Coordinate‑independent frequency embeddings (one for the initial layers, one for the remaining layers) perform comparably when chosen sufficiently large.
	\item[(\textbf{O6})] Increasing regularization weights generally improves results, but excessively large values can stall optimization.
\end{itemize}

Based on these observations, we adopted the following guidelines. Motivated by (\textbf{O2}), we fix
the network architecture to three hidden layers with $256$ neurons for both $m$ and $S$. Following
(\textbf{O4})-(\textbf{O5}), we use coordinate‑independent frequency embeddings chosen sufficiently
large. For each coordinate direction, we fix the same batch sizes for the continuous and discrete
domains by selecting the largest values allowed by memory (keeping the temporal batch size small).
Guided by (\textbf{O6}), we determine regularization weights through a short warm‑up procedure of
$500$ iterations: we enable one regularization term at a time, gradually increase its weight until
optimization stalls, and then choose a slightly smaller value. We add regularizers in the order
temporal TV $\rightarrow$ spatial TV  $\rightarrow$ coil regularization; the influence of this order
was not investigated.

These guidelines reduce the effective number of tunable hyperparameters to the number of modes and
the learning rate---only $2(n+1)$ parameters. While this strategy does not guarantee optimality, it
provides a practical and computationally efficient approach that avoids large‑scale hyperparameter
searches. For datasets $(i)$ and $(ii)$, we pick one specific scan (subject) on which we perform
this simplified hyperparameter search, and then use the found hyperparameters on all other subjects
in the datasets. The final hyperparameter settings for the three datasets are summarized in Table
\ref{table:final_params}. 

\begin{table}
\centering
\scalebox{0.95}{
    \begin{tabular}[ht!]{cccccccc}
        \toprule 
        \textbf{Dataset} & $\left(N_{t}, N_{x_{1}}, \ldots, N_{x_{n}} \right)$ & $\left(M_{x_{1}}, \ldots, M_{x_{n}} \right)$ & 
         \textbf{Memory (GPU)} & \textbf{Rec. time (min)} \\[1ex]
        \midrule
         $(i)$  & $(64, 96, 84)$ & $(64,48)$ & $800$MB-$1200$MB & $3$-$5$ \\ 
         $(ii)$ & $(84,96,64,64)$ & $(64,48,48)$ & 12GB & $45$ \\
         $(iii)$ & $(48,84,96,64)$ & $(48,64,48)$ & 9GB & $15$ \\
        \midrule
        \bottomrule				
    \end{tabular}}
    \caption{
        \label{table:final_params} Number of modes $\left(N_{t}, N_{x_{1}}, \ldots, N_{x_{n}}
        \right)$ and $\left(M_{x_{1}}, \ldots, M_{x_{n}} \right)$ used for the NFE expansions of the
        magnetization and coil sensitivities, respectively. GPU memory usage during optimization and
        approximate reconstruction times are also reported.  The networks use a fixed backbone with
        three hidden layers of $256$ neurons each.
    }     
\end{table}

\paragraph{Time-resolved NFE reconstructions}
For $(ii)$, we reconstruct directly from the raw unbinned $k$-space readouts, treating each readout
as an individual temporal observation; no temporal binning or grouping is applied. Following
reconstruction, each of the six pressure cycles is partitioned into nine phases corresponding to the
pressure levels for which fully sampled reference images exist. For quantitative evaluation, each
reconstructed phase is compared with its associated ground truth image using SSIM and PSNR.
Specifically, within each phase we evaluate the NFE at all observation times (readout moments
assigned to that phase) and select the time point whose reconstruction is most similar to the static
ground truth image, as measured by SSIM. SSIM and PSNR are then reported at this selected time
point. Repeating this procedure for nine phases across six pressure cycles yields a total of $54$
comparisons per subject. Because the reference images originate from a separate scan, residual
motion between scans is expected; therefore, each reconstructed phase is first rigidly registered to
the ground truth image using \textsc{ELASTIX} \cite{klein2009elastix, shamonin2014fast} prior to
metric computation.

\subsection{Results}
In this section we assess the performance of the proposed method for the three different applications. 

\subsubsection{$(i)$ Dynamic $2$D CMR dataset}
\begin{figure}[tb!]
	\centering
    	\includegraphics[width=0.87\linewidth,clip,trim=0 0 0 0]{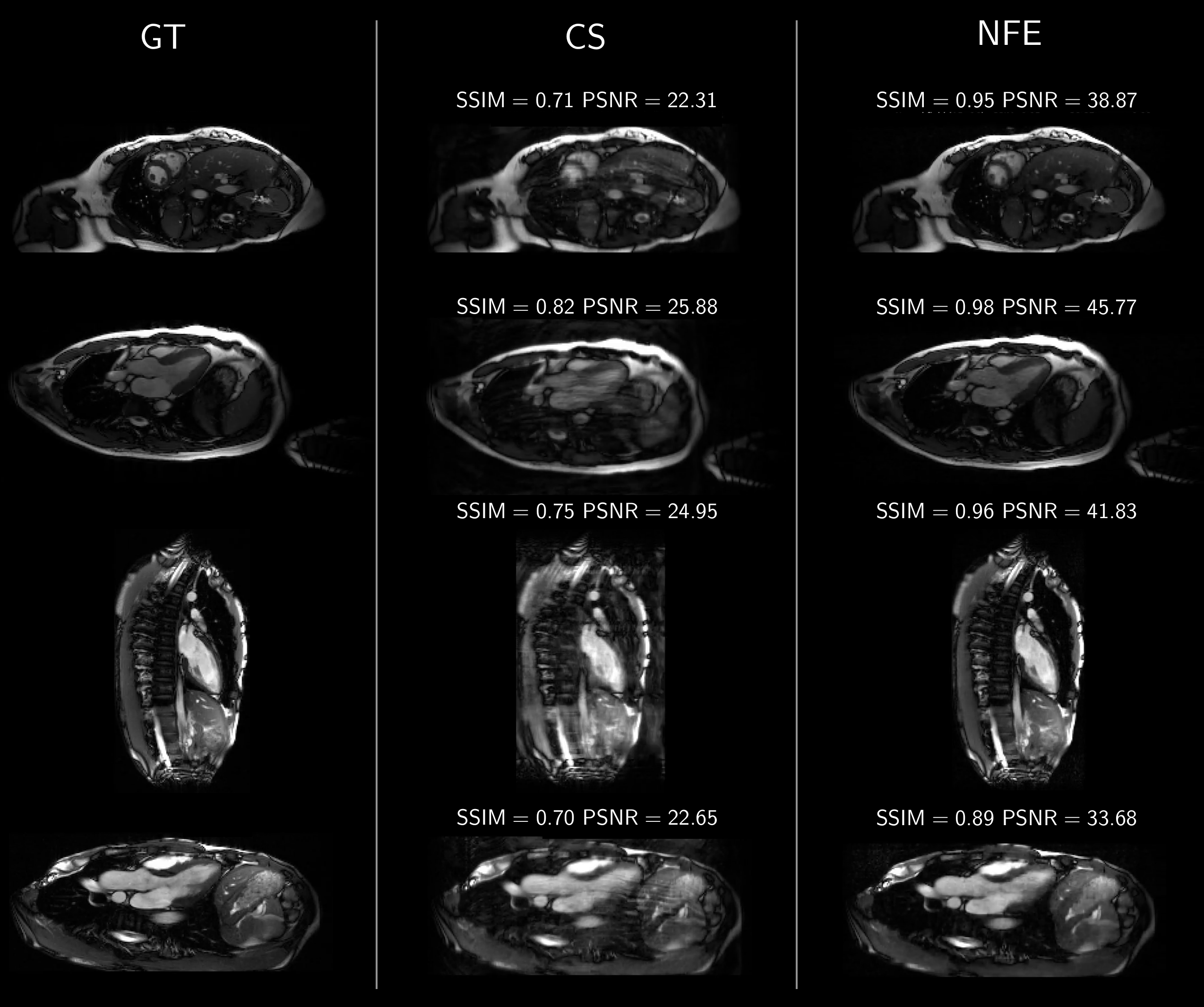}
  	\caption{
        Dynamic $2$D reconstruction of selected OCMR slices at $\text{AF}=16$. Columns show ground
        truth (GT), CS, and NFE reconstructions. Rows correspond to different slices at a single
        time frame: the first two rows show $3$T scans and the last two rows $1.5$T scans.
    }
 	 \label{fig:ocmr_fs_results}
\end{figure}

\begin{figure}[tb!]
	\centering
    	\includegraphics[width=\linewidth,clip,trim=0 0 0 0]{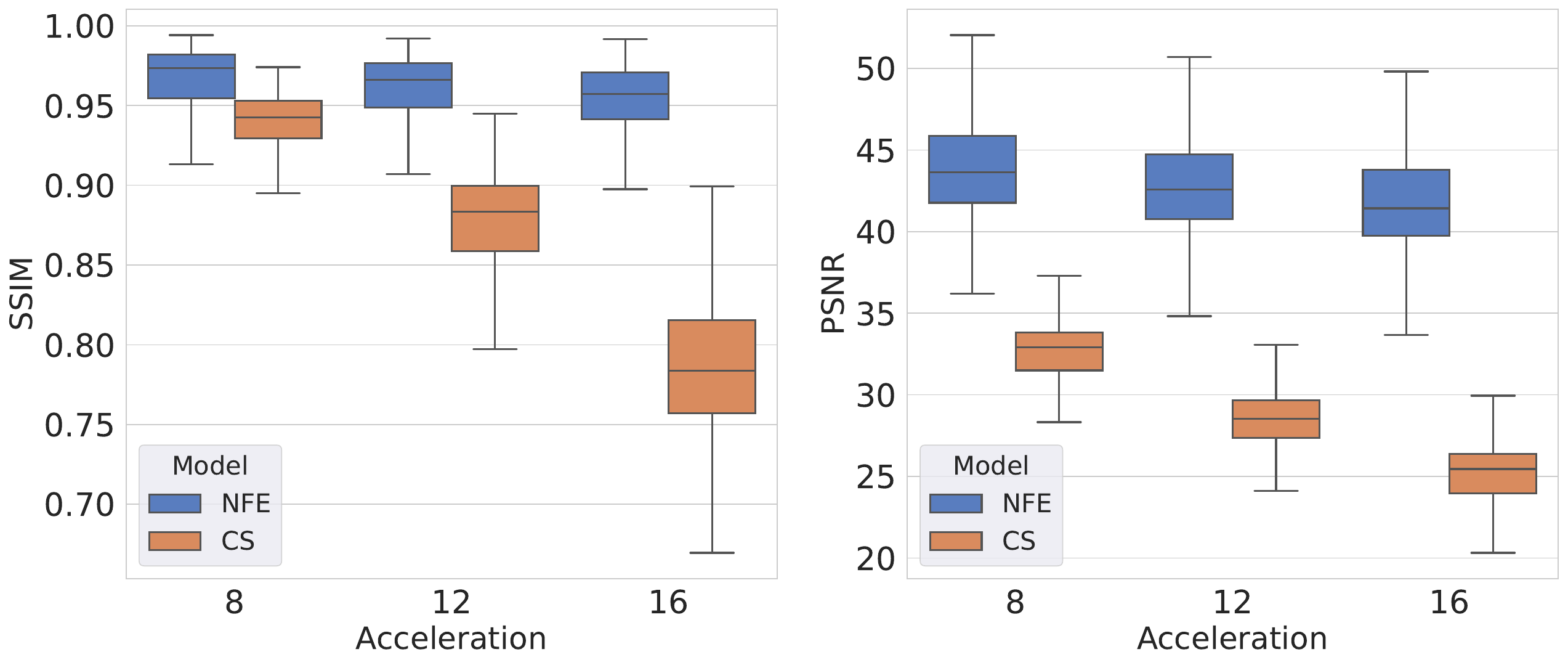}
    	\caption{
            Quantitative evaluation of the dynamic $2$D reconstructions. SSIM and PSNR are computed
            for NFE and CS reconstructions across acceleration factors $\{8,12,16\}$, evaluated over
            all slices and time frames.
        }
    	\label{fig:boxplot_ocmr}
\end{figure}

Figure \ref{fig:ocmr_fs_results} presents qualitative comparisons for selected slices at
acceleration factor (AF) $16$. Temporal dynamics across all acceleration factors are shown in
Supplementary Video S1 (S1a-S1d). At $\text{AF} = 8$, CS-based reconstructions exhibit pronounced
flickering and streaking artifacts when frames are viewed sequentially and fine cardiac structures
appear blurred. These artifacts intensify with higher AFs, rendering CS reconstructions visually
unreliable at AF $12$ and $16$. This degradation indicates that conventional compressed sensing
struggles to maintain temporal consistency and anatomical fidelity under aggressive undersampling,
even with strong regularization. The statistics in Figure \ref{fig:boxplot_ocmr} corroborate these
findings. 

In contrast, NFE‑based reconstructions maintain high visual quality across all AFs. No flickering or
streaking artifacts are observed. At $\text{AF}=8$, only a slight loss of very fine detail is
noticeable, becoming marginally more pronounced at higher AFs without affecting overall
interpretability. The weakest NFE performance occurs in a small number of challenging $1.5$T slices,
where stronger temporal regularization would likely have been beneficial. However, we intentionally
retained a single set of hyperparameters---tuned on a representative $3$T slice---to demonstrate
robustness and broad applicability across all $165$ subjects. 

\subsubsection{$(ii)$ Dynamic $3$D MRI dataset}
\begin{figure}[tb!]
	\centering
    	\includegraphics[width=\linewidth,clip,trim=0 0 0 0]{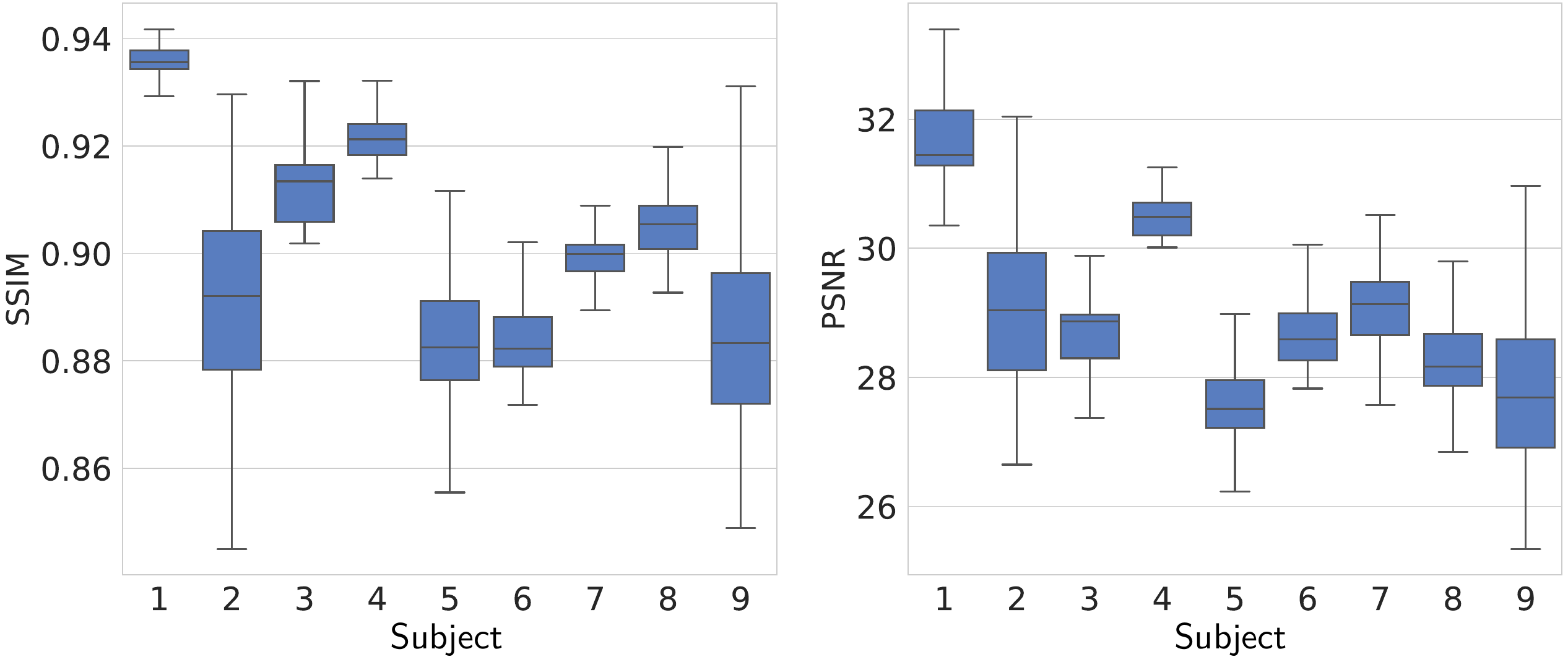}
    	\caption{
            Quantitative evaluation on the dynamic $3$D thigh dataset. For each subject, the six
            pressure cycles are partitioned into nine phases corresponding to the pressure levels
            with fully sampled reference images. For each phase, the neural field is evaluated at
            all associated readout times and the time point best matching the static reference is
            selected. SSIM and PSNR are reported for the resulting $54$ reference comparisons per
            subject.
        }
    	\label{fig:boxplot_thigh}
\end{figure}

\begin{figure}[tb!]
	\centering
	  \begin{subfigure}{\linewidth}
    		\includegraphics[width=\linewidth,clip,trim=0 0 0 0]{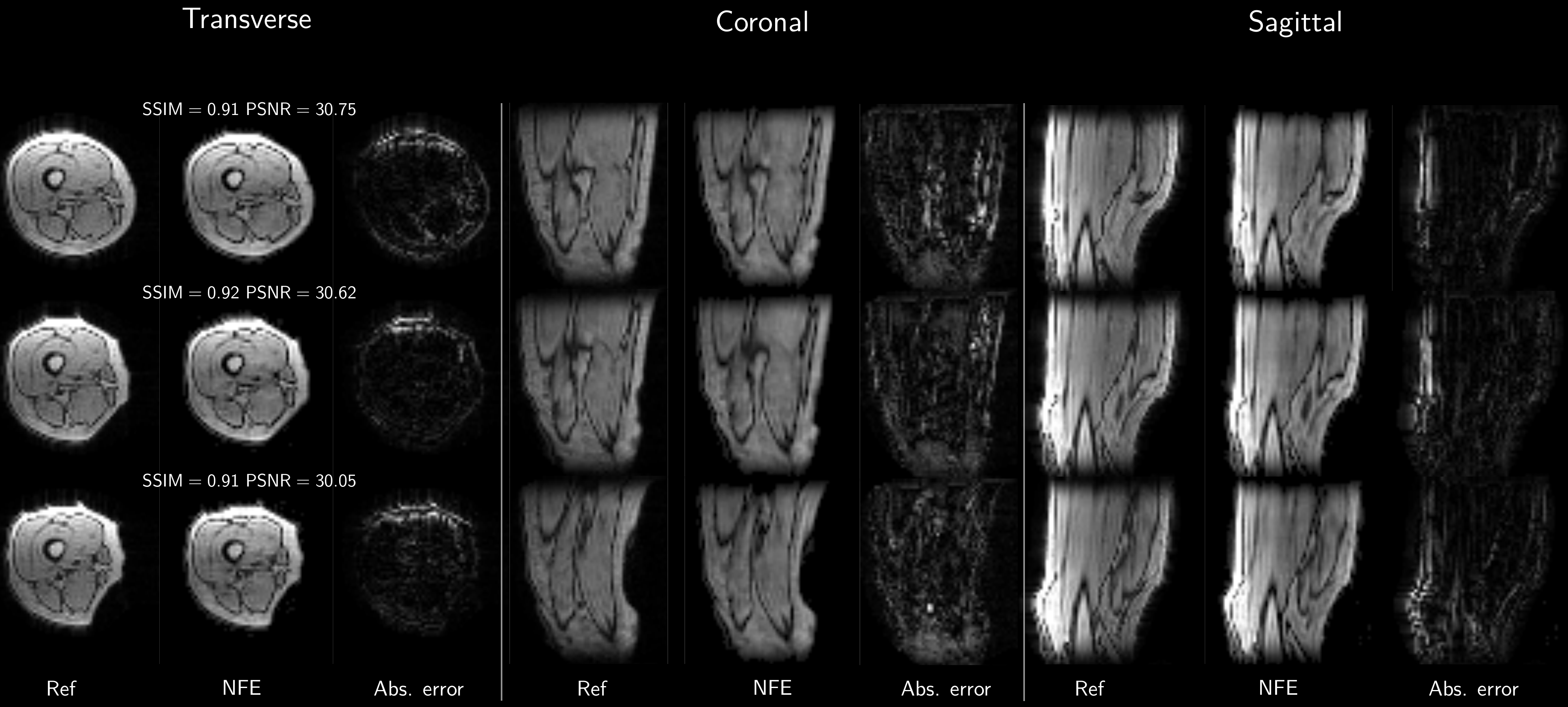}
    		\caption{Subject $1$}
		\label{fig:vol1}
	\end{subfigure} \\[2ex]
	\begin{subfigure}{\linewidth}
    		\includegraphics[width=\linewidth,clip,trim=0 0 0 0]{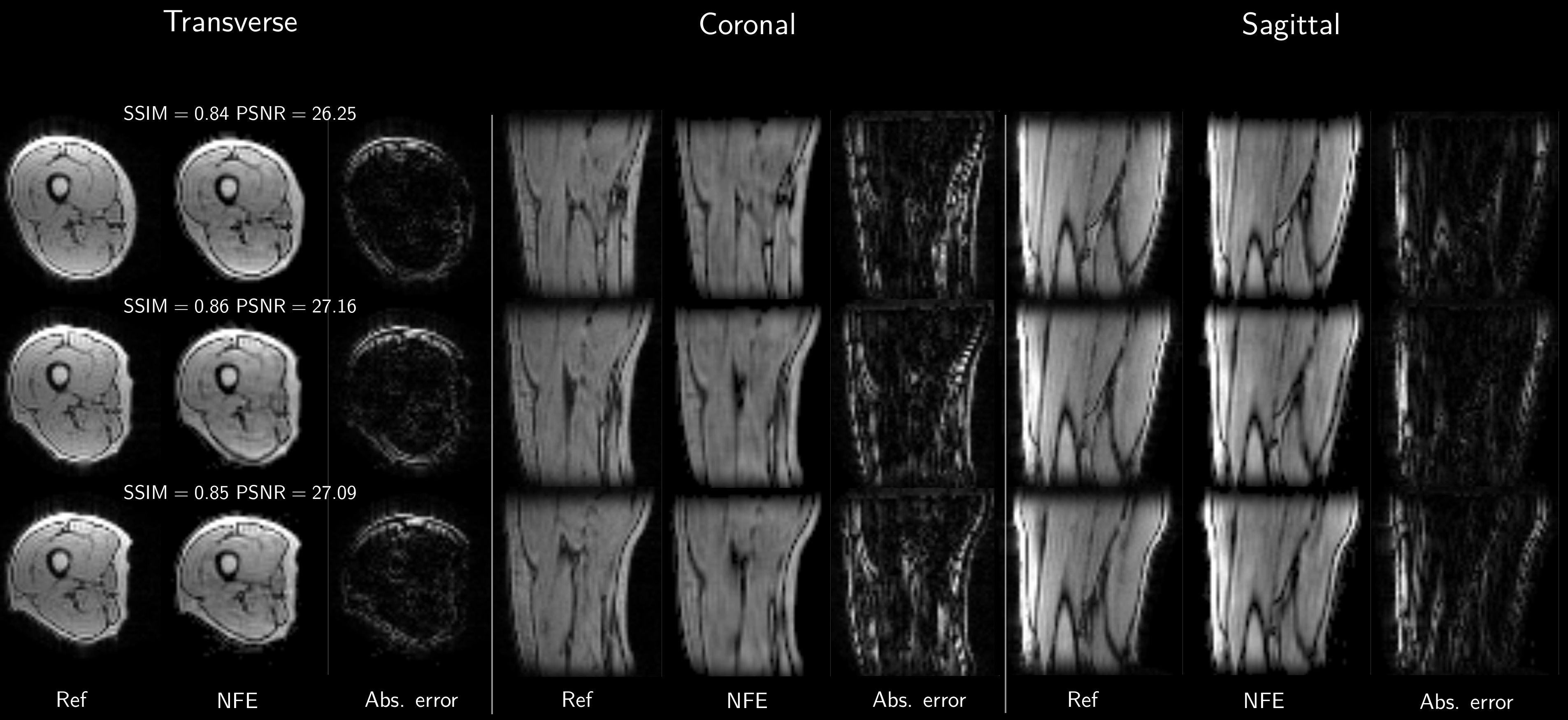}
    		\caption{Subject $2$}
		\label{fig:vol2}
	\end{subfigure}
	\caption{
        Dynamic $3$D reconstruction of two subjects in the thigh dataset for three representative
        phases: no deformation (first row), minor deformation (second row), and maximal deformation
        (third row) within the pressure cycle. Columns are grouped by anatomical plane, showing
        transverse, coronal, and sagittal views. Within each group, the first image is the fully
        sampled reference (Ref), the second is the NFE reconstruction, and the third is the absolute
        difference. Subjects $1$ and $2$ represent the best‑ and worst‑performing cases,
        respectively.
    }
    \label{fig:thigh_comparisons}
\end{figure}

The time‑resolved NFE reconstructions exhibit some slight noise, which is expected in the absence of
temporal binning or clustering, as shown in Supplementary Video S2 (S2a-S2b). A small number of
outliers appear for subjects $2$, $5$ and $9$, where SSIM drops noticeably, as shown in Figure
\ref{fig:boxplot_thigh}. In Figure \ref{fig:thigh_comparisons} we present the best‑ and
worst‑performing cases. The discrepancies originate primarily along the outer fat–muscle interface,
where the NFE reconstructions display sharper and thicker boundaries---likely a consequence of
relatively strong spatial TV regularization. These localized boundary‑intensity differences lead to
reduced SSIM in the outlier cases and, more importantly, systematically lower PSNR across all
comparisons, since PSNR is highly sensitive to small intensity mismatches.

Crucially, these findings must be interpreted in the context of the validation setup: we compare
time‑resolved reconstructions against static fully sampled reference images obtained from a separate
scan. Residual motion, deformation, and global intensity drift between scans cannot be fully
corrected by rigid registration. As a result, very high similarity measures are not expected in this
setting. Despite these limitations, the boxplots in Figure \ref{fig:boxplot_thigh} and visual
comparisons confirm that the NFE reconstructions remain anatomically consistent and temporally
coherent across all subjects.

\subsubsection{$(iii)$ Dynamic $3$D CMR}
Qualitative comparisons for selected slices across the three orthogonal directions are shown in
Figure \ref{fig:ohio_us}. Animations of the reconstructed cardiac cycle are provided in
Supplementary Video S3 (S3a-S3b). As in the $2$D case $(i)$, CS-based reconstructions exhibit severe
flickering, streaking artifacts, and blurring of fine structures, resulting in an overall noisy
appearance.  In contrast, NFE-based reconstructions retain anatomical detail more effectively,
exhibit substantially less noise, and remain free of flickering artifacts when frames are viewed
sequentially. 


\section{Discussion and Conclusion}
\begin{figure}[t!]
	\centering
    	\includegraphics[width=0.75\linewidth,clip,trim=0 0 0 0]{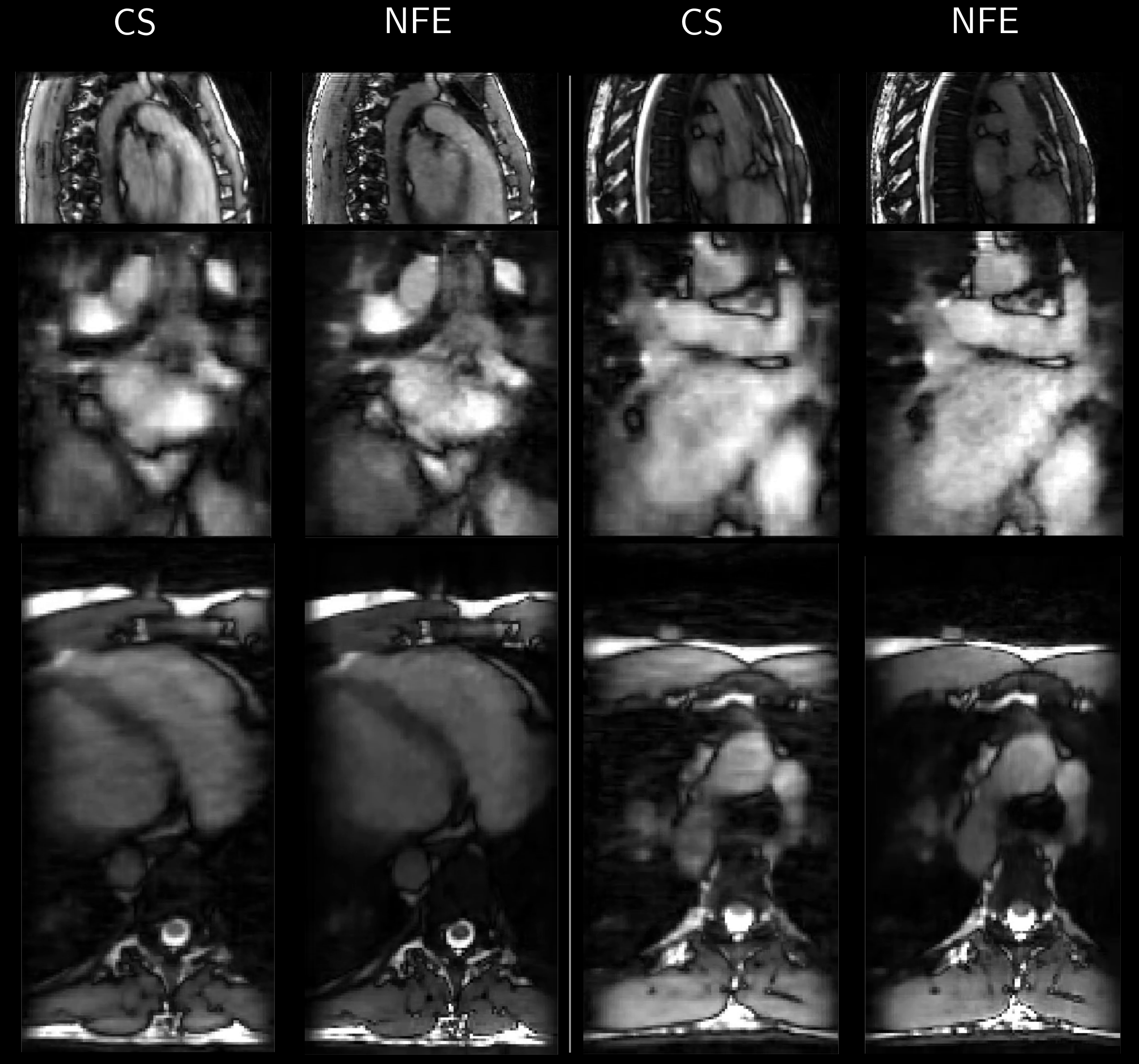}
  \caption{
    Dynamic $3$D CMR reconstruction at  $\text{AF}=16$. Two representative temporal frames are shown
    (left and right column groups). Within each group, CS reconstructions are compared against NFE
    reconstructions. Rows correspond to slices in the three orthogonal directions.
  }
  \label{fig:ohio_us}
\end{figure}
In this work we demonstrated the effectiveness of continuous representations for MRI reconstruction
through neural field expansions (NFEs). By exploiting a tensor product structure together with
stochastic optimization, our framework substantially reduces memory requirements, enabling
high‑acceleration dynamic $2$D and $3$D MRI reconstruction. In particular, we achieved a $31 \times$
reduction in our dynamic $2$D example, and in the dynamic 3D cases reduced memory usage from being
infeasible on a $48$GB GPU to only needing $9$-$12$GB. Across both quantitative and qualitative
evaluations, NFEs consistently outperform state‑of‑the‑art finely tuned CS methods, preserving fine
spatial detail and dynamic motion even under aggressive undersampling. Furthermore, NFEs can
reconstruct directly from fully unbinned raw $k$-space readouts without any temporal grouping or
preprocessing---an operation that is computationally infeasible with standard CS pipelines. In our
second application, this allowed us to target a temporal resolution of $5.4$ ms (equal to the TR).
Our method also does not require fully sampled central $k$-space data for coil calibration, offering
greater flexibility in sampling design and supporting higher acceleration rates, particularly in
dynamic $3$D MRI.

While NFEs are robust to high undersampling, extremely sparse sampling can still limit their ability
to capture low‑amplitude temporal motion. Nonetheless, across all acceleration factors evaluated,
NFEs preserve anatomical structures and motion more faithfully than CS methods, whose reconstruction
quality deteriorates sharply under similar conditions. These trends are reflected in the
quantitative metrics, which show consistently higher SSIM and PSNR for NFE‑based reconstructions.
This consistency suggests that NFEs scale well to higher‑dimensional dynamic MRI, preserving spatial
fidelity and temporal coherence under conditions that challenge CS‑based approaches.

A remaining challenge is the relatively large number of hyperparameters in the NFE framework.
Although NFEs are less sensitive to these choices than traditional CS methods, the parameter space
remains substantial. To keep hyperparameter tuning tractable, we adopted heuristic guidelines
informed by an exhaustive search on a representative subset of the OCMR dataset. This choice
sacrifices some performance, but provides a practical path toward reproducible and scalable
applications. Computational costs, however, could likely be reduced further by decreasing the number
of modes or selecting more optimal batch sizes to improve convergence speed.

Our evaluation on dynamic $3$D CMR data was limited to a single moderate‑quality scan. Here we
cannot claim optimal performance of either method (CS and NFEs), as ground truth is unavailable and
exhaustive hyperparameter tuning is infeasible without a ground truth. A broader assessment on more
challenging, higher‑quality datasets is needed to fully characterize the strengths and limitations
of both NFEs and CS. While the present study on dynamic $3$D CMR relied on temporally binned cardiac
data, future work will target direct optimization on unbinned dynamic $3$D acquisitions. Progress in
this direction is currently constrained by the limited availability of publicly released dynamic
$3$D CMR datasets, highlighting the need for increased data sharing in the community.

Finally, although we initially drew an analogy between NFEs and generalized Fourier series, this
connection remains formal: NFE modes are neither orthonormal nor guaranteed to be linearly
independent. As a result, mode contributions cannot be interpreted directly from coefficient
magnitudes, which limits interpretability even though reconstruction quality is unaffected. Future
research will investigate orthogonalization strategies during training---for example, via modified
Gram–Schmidt procedures---to reduce redundancy, improve interpretability, and place NFEs on firmer
mathematical foundations.

\bibliographystyle{plain}
\bibliography{bibliography} 
\nocite{*}

\end{document}